\documentclass[preprint,superscriptaddress,preprintnumbers,pra,showpacs]{revtex4}
\usepackage{amsmath}
\usepackage{graphicx}

\newcommand\be{\begin{equation}}
\newcommand\ee{\end{equation}}

\begin{document}

\title{Nonlinear optical anisotropy due to freed electrons}

\author{J. Andreasen}
\affiliation{
  College of Optical Sciences, University of Arizona,
  Tucson, AZ 85721
}
\author{E. M. Wright}
\affiliation{
  College of Optical Sciences, University of Arizona,
  Tucson, AZ 85721
}
\author{M. Kolesik}
\affiliation{
  College of Optical Sciences, University of Arizona,
  Tucson, AZ 85721
}
\affiliation{
  Department of Physics, Constantine the Philosopher University,
  Nitra, Slovakia
}

\date{\today}

\begin{abstract}
We present a computational study of the pump-probe response of a single atom to assess any microscopic nonlinear 
optical anisotropy due to freed electrons for non-resonant optical excitation.  
Using simulations of the Schr\"odinger equation for an atom exposed to a strong, short-duration linearly polarized 
pump, we calculate the induced dipole moment along probe directions
parallel and perpendicular to the pump polarization.
Our simulations show birefringence ratios of approximately 0.78--0.9
on the timescale of tens of femtoseconds following the excitation pulse.
\end{abstract}

\maketitle

\section{Introduction}

Isotropic dielectric media, such as gases or cubic crystals, can exhibit nonlinear 
birefringence for incident field frequencies in the transparency region \cite{boyd03}.  
This nonlinear optical anisotropy (NOA) is revealed in pump-probe schemes in which a linearly polarized pump field 
is incident on the medium along with a co-propagating weak probe field polarized either parallel or perpendicular 
to the pump.  
For the case of a Kerr nonlinear medium, this yields a difference in refractive index experienced by the 
orthogonally polarized probes that is proportional to pump intensity. 
In other words, the orthogonally polarized probes induce different dipole moments along their 
respective directions in the presence of the pump.

Various pump-probe schemes are at the heart of recent approaches to measure the nonlinear optical properties 
and filamentation dynamics of atmospheric pressure gases, such as air or argon 
\cite{loriotOE09,wahlstrandOL11,odhner_ionization-grating-induced_2012,odhner_direct_2012}. 
If the pump pulse is sufficiently intense to generate freed electrons via multi-photon or tunneling ionization, 
this raises the issue of whether these electrons also contribute to the NOA.  
Here, we use the term freed electrons to describe those that are no longer bound, but may still be correlated 
with the parent ion, as opposed to strictly free electrons.  
Physically, the linearly polarized pump pulse drives the freed electrons along the polarization direction, 
thereby creating an anisotropic momentum and spatial distribution for the electrons.  
However, it is a common assumption that freed electrons exhibit an isotropic optical response. 
This is true if the freed electrons and associated ions have had ample time to undergo Coulomb 
scattering and produce a collective plasma whose momentum distribution is isotropic.  
From simulations of the quantum Boltzmann equation for an atomic gas,
it has been established \cite{pasMolKoc2011} that anisotropic features of the momentum distribution 
survive for a time of order $t_{a}\simeq 1/\omega_{pl}$, given by the inverse plasma frequency.
A collective isotropic plasma appears for longer times $t_{cp}$, with $t_a\simeq 500$ fs and $t_{cp}$ a few 
picoseconds for conditions typical of femtosecond filamentation in atmospheric pressure gases.  
Thus, for pulse durations greater than $t_{cp}$, freed electrons should yield an isotropic optical response.  
However, for pulse durations less than $t_{a}$, we expect that any anisotropy due to freed electrons will 
persist and may lead to a NOA.  
Photo-induced birefringence in semiconductors due to single-photon ionization provides an example of a NOA due to 
freed electrons for optical fields \cite{WanAllLen2007}.

It has been recently shown \cite{pasenowPC} that an anisotropic electron momentum distribution 
can reveal itself using a THz probe.  
In this paper, we present a computational study of the pump-probe response of a single atom, in order to assess any 
microscopic NOA due to freed electrons for optical fields. 
Essentially, we inquire if the freed electrons retain a ``memory'' of the linear polarization of the pump pulse that 
generated them.
We restrict simulated pulse durations to less than $t_a$, so that anisotropy of the electron distribution due to the 
pump pulse will be minimally degraded by electron-electron interactions, and so that a single atom approach is 
sufficient.
In particular, using simulations of the single particle Schr\"odinger equation,
the induced dipole moment along two probe directions is calculated to assess any NOA.  
Two pump models are examined: a delta-function pump of zero temporal duration, and a realistic pump with 
non-zero duration. 
In both cases, we find that a sizeable anisotropic nonlinear atomic response can occur due to freed electrons. 
Our hope is that these computational results provide impetus for future experimental and theoretical studies of 
NOA due to freed electrons on ultrashort time scales.  
In addition to being of fundamental interest, our results may have implications for the interpretation of ultrashort
pump-probe experiments aimed at measuring nonlinear optical properties in the presence of freed electrons 
(e.g. \cite{loriotOE09,wahlstrandOL11}) since they rely on the assumption of an isotropic freed-electron response. 
Similarly, ``plasma'' isotropy is invariably assumed in computer aided modeling of femtosecond 
filamentation~\cite{Guide11}, so the current model may require corrections.

\section{Numerical model}

Our starting point is the three-dimensional (3D) Schr\"odinger equation in atomic units (a.u.) for the wave function 
$\Psi(\vec{r},t)$ of an electron under the actions of the Coulomb potential due to the parent ion and the pump and 
probe fields
\begin{equation}\label{SWE}
  i\frac{\partial\Psi}{\partial t} = \left[-\frac{1}{2}\nabla^2-\frac{1}{r} +V(\vec r,t)\right]\Psi  ,
\end{equation}
where $V(\vec r,t)=\vec r\cdot[\vec E_P(t)+\vec E_{pr}(t)]=[V_P+V_{pr}]$ is the interaction energy between the atom 
and pump ($P$) and probe ($pr$) fields in the dipole approximation.  
In the absence of pump and probe fields, the spherically symmetric ground state wave function is 
$\Psi_g(\vec r,t)=\psi_g(r)\exp(-i\epsilon_g t)$, with 
$\psi_g(r)={\cal N}\exp(-r)$, $\cal N$ a normalization constant, and $\epsilon_g$ the ground state energy.  

We take the pump field to be linearly polarized, $\vec E_P(t)=\vec e_z E_P(t)$, with $\vec e_z$ the unit vector 
along the $z$-axis. 
Then, employing cylindrical coordinates $\vec r=(\rho,\theta,z)$, the cylindrically symmetric solution 
$\Psi(\vec r,t)=\Psi_P(\rho,z,t)$ of Eq. (\ref{SWE}) in the presence of the pump beam alone evolves as
$i{\partial \Psi_P/\partial t}=\hat H_P\Psi_P$ with the Hamiltonian
\begin{equation}\label{HP}
  \hat H_P = -\frac{1}{2}\left (\frac{1}{\rho}\frac{\partial}{\partial \rho}\rho\frac{\partial}{\partial \rho}+\frac{\partial^2}{\partial z^2} \right )-\frac{1}{r} + zE_P(t)  .
\end{equation}
The probe is assumed much weaker than the pump, which makes $\Psi_P$ the zeroth-order wave function, and
the atom-probe interaction is treated to first-order in perturbation theory.  
The probe field is written as $\vec E_{pr}(t)=\vec e_m E_{pr}(t)$, where $m=0$ corresponds to the probe polarized 
along the $z$-axis, $\vec e_0=\vec e_z$, and $m=1$ to the probe polarized along the $x$-axis, $\vec e_1=\vec e_x$. 
We express $V_{pr}=\xi_m\cos(m\theta)E_{pr}(t)$, with $\xi_0=z$ and $\xi_1=\rho$.  
Writing the wave function as $\Psi(\vec r,t)=\Psi_P(\rho,z,t)+\psi_m(\rho,z,t)\cos(m\theta)$, 
the Schr\"odinger equation for the first-order correction due to atom-probe interactions is
\begin{equation}\label{psim}
  i\frac{\partial\psi_m}{\partial t} = \left [\hat H_P + \frac{m^2}{2\rho^2} \right ]\psi_m + \xi_mE_{pr}(t)\Psi_P(\rho,z,t) ,
\end{equation}
with the initial condition $\psi_m(\rho,z,0)=0$.  
We have built into this equation, the same probe field $E_{pr}(t)$ applied along both axes.

Our approach involves generating numerical solutions for $\Psi_P$ and $\psi_{0,1}$,
which are used to calculate the light-induced atomic dipole moments.  
In particular, the dipole moment due to the pump alone
\begin{equation}\label{dP}
  d_P(t) = 2\pi \int_0^\infty d\rho\rho \int_{-\infty}^\infty dz\Psi_P^*z\Psi_P  ,
\end{equation}
with the resulting dipole moment directed along the $z$-axis. 
The probe-induced changes to the dipole moment
\begin{equation}\label{dm}
  d_m(t)=\frac{4\pi}{(1+m)}\mbox{Re}\left[\int_0^\infty d\rho\rho\int_{-\infty}^\infty dz\Psi_P^*\xi_m \psi_m\right],
\end{equation}
where $d_m(t)$ is directed along the $z$-axis ($x$-axis) for $m=0$ ($m=1$)
and corresponds to the raw data that we extract from the simulations.
Furthermore, we also calculate the probe-induced changes to the dipole moment in the absence of the pump beam
$d_m^{(g)}(t)$, which are obtained using Eq. (\ref{dm}) with $\Psi_P\rightarrow\Psi_g$.  
The nonlinear change in the dipole moment is then given by $\delta d_m(t)=[d_m(t)-d_m^{(g)}(t)]$.
We shall discuss further processing of the data as we proceed.

Turning now briefly to some details of the numerical approach, the Schr\"odinger equations for $\Psi_P$ and 
$\psi_{0,1}$ are converted to systems of ordinary differential equations on a discrete grid in the two spatial 
coordinates $(\rho,z)$. 
The Coulomb potential is discretized by asymptotic behavior correspondence to ensure accuracy and improve 
computational efficiency \cite{gordonPRA06},
and we begin the simulation in the ground state of the discretized system of equations.
We perform a convergence study of measured observables to ensure that both radial and longitudinal grid sizes are
sufficient to accommodate the expanding, pump-driven wave function. 
The results shown here use a grid resolution of 0.3~a.u. with a computational box size up to 9000~a.u by 250~a.u.
in the $z$ and $\rho$ directions, respectively.
Because we calculate and compare the induced dipole moments in different directions, our approach here is to avoid 
using absorbing boundaries which could affect results. 
Rather, we choose to restrict our simulations to times short enough such that a convergence study shows absence of 
boundary-related artifacts.

\section{Results}

\begin{figure}
  \centering
  \includegraphics[width=8.0cm]{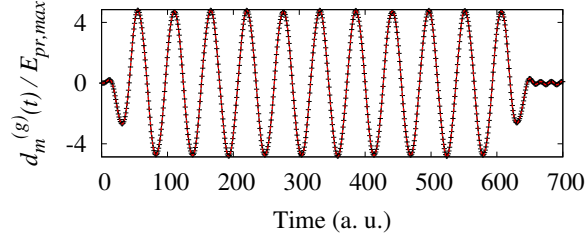}
  \caption{\label{fig:linear} (Color online)
    Scaled probe-induced dipole moment $(d_m^{(g)}(t)/E_{pr,max})$ versus time, 
    for the cases (black crosses) $m=0$ ($z$-polarized probe)
    and (red line) $m=1$ ($x$-polarized probe).
    This simulation verifies that isotropy is properly preserved in absence of the pump.
  }
\end{figure}

The goal of these calculations is to detect any anisotropy that may exist in the nonlinear response caused by a
linearly polarized excitation.
Thus, we verify that the model presented properly preserves isotropic properties where appropriate.
To test the utility of the perturbative equations (\ref{psim}), we solve for the probe-induced dipole moments 
in the absence of the pump beam for off-resonant probe fields.
Figure \ref{fig:linear} shows the scaled, probe-induced dipole moments $(d_m^{(g)}(t)/E_{pr,max})$ versus time for
the case $m=0$ ($z$-polarized) and the case $m=1$ ($x$-polarized).
A probe field $E_{pr}(t)$ is used with maximum amplitude $E_{pr,max}$ and a temporal profile matching $d_m^{(g)}(t)$
shown in the figure.  
We see that $d_0^{(g)}(t)=d_1^{(g)}(t)$, meaning the linear probe response is isotropic, as expected.
Furthermore, we find that $d_m^{(g)}(t)=\alpha E_{pr}(t)$, with $\alpha\gtrsim 4.5$, very close to the 
analytic result for the off-resonant linear polarizability of the atom.  
Thus, we have validation of the approach for the linear optical response.

We now turn to simulations of the pump-probe response.  
The first model considered is for a delta-function pump pulse of the form $E_P(t)=F\delta(t)$.
Despite the extreme nature of this pulse, namely that it has zero temporal duration and hence contains all 
frequencies in equal proportion, we study this model for three main reasons.
First, following the $z$-polarized pump pulse, an atom initially in its ground state becomes 
$\Psi_P(\vec r,t=0+)=\psi_g(r)\exp(-iFz)$ \cite{girardeauPRA92}, which eliminates the need for a numerical solution 
to model the pump modification of the wave function.  
Second, this provides a simple model for the case when the pump and probe are temporally separated.  
In this limit, the probe experiences the atomic state left behind by the pump pulse but with no NOA due to the Kerr 
effect, since the pump has passed.  
Finally, we are optimistic that the simplifications this model affords will eventually yield analytic insights into 
the physics underlying the NOA.

\begin{figure}
  \centering
  \includegraphics[width=8.0cm]{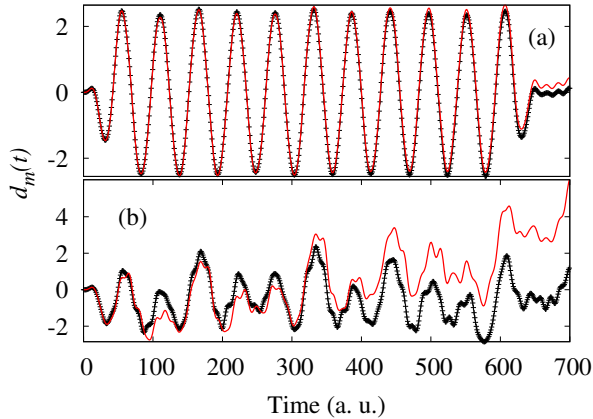}
  \caption{\label{fig:deltaFull} (Color online)
    Calculated probe-induced dipole moments (red line) $d_0(t)$ and (black crosses) $d_1(t)$ for a delta-function 
    pump with (a) $F=0.041$, and (b) $F=0.17$. 
    The probe has a center wavelength of $\lambda_{pr}=800$ nm and has a temporal profile profile shown by the 
    blue line in Fig. \ref{fig:deltaFiltered}(a).
  }
\end{figure}

\begin{figure}
  \centering
  \includegraphics[width=8.0cm]{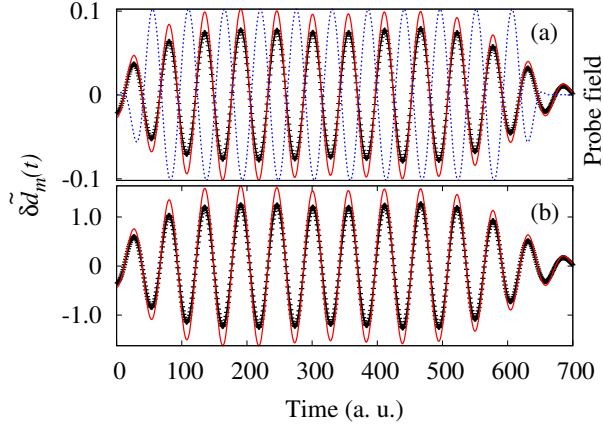}
  \caption{\label{fig:deltaFiltered} (Color online)
    Filtered nonlinear changes in dipole moment for (red line) $\delta\tilde d_0(t)$ and 
    (black crosses) $\delta\tilde d_1(t)$ for a delta-function pump with (a) $F=0.041$ and (b) $F=0.17$. 
    The blue dashed line in (a) shows the temporal profile of the probe in arb. units.
  }
\end{figure}

Figure \ref{fig:deltaFull} shows the calculated raw data in the form of the probe-induced dipole moments $d_{0,1}(t)$
for a delta-function pump pulse with $F=0.041$ and $F=0.17$. 
In each case, the probe pulse has a center wavelength of $\lambda_{pr}=800$ nm and a temporal profile shown
in Fig. \ref{fig:deltaFiltered}(a). 
Based on the results from Ref. \cite{girardeauPRA92}, the ionization probability is $0.1$\% for $F=0.041$ and 
$1.8$\% for $F=0.17$.
In each case the pump leaves behind the probability of a freed electron that can modify the probe.  
We note that for the relatively weak pump of $F=0.041$ in Fig. \ref{fig:deltaFull}(a), 
the temporal profile of the dipole moments $d_{0,1}(t)$ closely follow the probe profile in 
Fig. \ref{fig:deltaFiltered}(a), indicating a weak nonlinear response,
whereas the temporal profiles of the dipole moments $d_{0,1}(t)$ for $F=0.17$ in Fig. \ref{fig:deltaFull}(b)
are clearly different and show a much stronger nonlinear effect.

We are primarily interested in the nonlinear change in dipole moment for frequencies in the vicinity of the probe 
frequency $\omega_{pr}$, and for this we employ the filtering procedure described in 
Ref. \cite{BroWriMol2012}.
Obtaining the nonlinear change in dipole moment $\delta d_m(\omega)$ from  $\delta d_m(t)$ via a Fourier transform,
a band-pass filter ${\cal F}(\omega-\omega_{pr})$ centered on $\omega_{pr}$ is applied to isolate the nonlinear 
response in the vicinity of the probe. 
The inverse Fourier transform yields the filtered nonlinear change in dipole moment $\delta\tilde d_m(t)$.  
The robustness of this filtering procedure has been tested by considering various Gaussian and super-Gaussian 
filters.
Figure \ref{fig:deltaFiltered} shows $\delta\tilde d_{0,1}(t)$ for $F=0.041$ and $F=0.17$.
There are several key features of these results.
First, despite the differences in the raw data in Figs. \ref{fig:deltaFull}(a) and \ref{fig:deltaFull}(b), 
the filtered nonlinear changes in dipole moment in Fig. \ref{fig:deltaFiltered} are very similar in temporal profile,
though they have different magnitudes.  
This highlights the importance of subtracting the probe-induced change in the absence of the pump. 
Second, $\delta\tilde d_{0,1}(t)$
are $180^\circ$ out of phase with the probe pulse [see Fig. \ref{fig:deltaFiltered}(a)], 
which is consistent with an optical response due to freed electrons.
Finally, for both examples in Fig. \ref{fig:deltaFiltered}, $\delta\tilde d_0(t)$ along the $z$-axis
is seen to be greater than $\delta\tilde d_1(t)$ along the $x$-axis, thereby demonstrating a NOA due to 
the freed electrons, a key result of this paper.  
There is no NOA due to the Kerr effect as the pump and probe do not overlap.

\begin{figure}
  \centering
  \includegraphics[width=8.0cm]{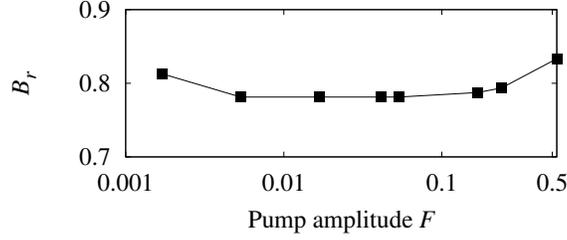}
  \caption{\label{fig:anisotropy}
    Birefringence ratio $B_r={\delta d_{1,peak} / \delta d_{0,peak}}$ as a function of the pump parameter
    $F$, with $\delta d_{m,peak}$ the peak value of $\delta d_{m}(\omega)$ near $\omega=\omega_{pr}$.
  }
\end{figure}

To further explore the NOA revealed above, Fig. \ref{fig:anisotropy} shows the birefringence ratio
$B_r={\delta d_{1,peak} / \delta d_{0,peak}}$ as a function of the pump parameter $F$,
with $\delta d_{m,peak}$ the peak value of $\delta d_{m}(\omega)$ near $\omega=\omega_{pr}$.
We checked that the results in Fig. \ref{fig:anisotropy} are largely unchanged by variations in the filtering 
procedure.
The calculation is limited at small values of $F$ ($< 0.001$) by numerical noise arising from the subtraction of the
relatively large probe-induced dipole moment in the absence of the pump. 
At large values of $F$ ($> 0.5$), the momentum supplied to the atomic wave function due to the pump pushes it 
strongly in one direction.
Therefore, the large zero-frequency component dominates the spectrum, making isolation of the nonlinear change to the
dipole moment in the vicinity of the probe frequency increasingly difficult.  
Over the range shown of $F$, the birefringence ratio varies between $B_r\approx 0.78-0.84$. 
To put this in context, using the results from Ref. \cite{girardeauPRA92}, the ionization probability is $0.1$\% for 
$F=0.041$ and $1.8$\% for $F=0.17$, so that these significant birefringence ratios arise for ionization 
probabilities that are not particularly large and are compatible with the conditions present during laser 
filamentation.

\begin{figure}
  \centering
  \includegraphics[width=8.5cm]{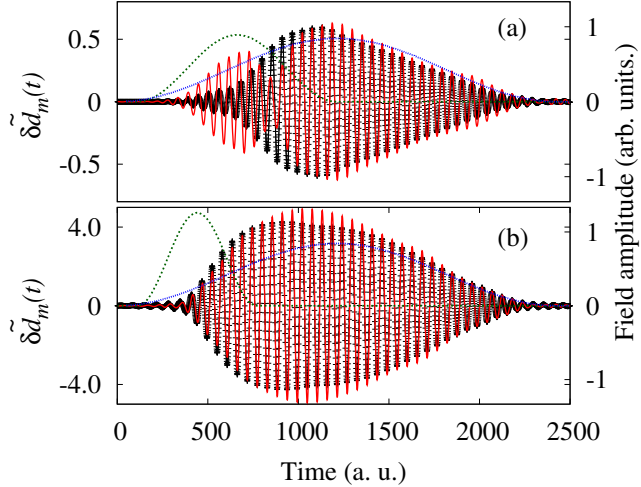} 
  \caption{\label{fig:long} (Color online)
    Filtered nonlinear changes in dipole moment for (red line) $\delta\tilde d_0(t)$ and 
    (black crosses) $\delta\tilde d_1(t)$.
    Also shown are the envelopes of the (blue dashed line) scaled-probe field and (green dotted line) pump field 
    with (a) intensity $10^{14}$ W/cm$^2$ and duration 13 fs, and
    (b) intensity $2\cdot 10^{14}$ W/cm$^2$ and duration 8 fs.
  }
\end{figure}

We next consider a more realistic model with a pump pulse of center wavelength $\lambda_P=800$ nm, peak intensity 
$10^{14}$ W/cm$^2$, and $13$ fs duration along with a weak probe pulse of center wavelength $\lambda_{pr}=400$ nm 
and duration $30$ fs.  
We choose the probe at the second-harmonic of the pump to aid in filtering out the change in dipole moment in the 
vicinity of the probe wavelength.  
The envelopes of the pump field and scaled probe field are shown in Fig. \ref{fig:long}(a), 
which are seen to overlap.  
We also note that the pulse durations are much shorter than the time scale $t_a\simeq 500$ fs, on 
which anisotropic features in the electron momentum distribution due to the pump pulse are expected to disappear. 

Figure \ref{fig:long}(a) shows the filtered nonlinear changes in dipole moment $\delta\tilde d_{m}(t)$ versus time 
with the probe field directed along the $z$-axis ($x$-axis) for $m=0$ ($m=1$).
For $t>1200$ a.u., after the pump pulse has passed, we again see that $\delta\tilde d_{0}(t)>\delta\tilde d_{1}(t)$,
indicating the presence of a NOA.  
Furthermore, the filtered nonlinear changes in dipole moment at these times are $180^\circ$ out of phase with 
the probe field so that the NOA is due to freed electrons.  
Inspection of the results for $t>1200$ a.u. yields a birefringence ratio $B_r\simeq 0.88$, closer to unity than 
for the delta-function excitation, but still significant.  
For $t<750$ a.u., it appears that the NOA is much more prominent with $B_r \approx 1/3$.  
During this time region the filtered nonlinear changes in dipole moment are in-phase with the probe field and the 
probability of ionization is small, indicating the observed NOA is dominantly due to the nonlinear 
Kerr effect \cite{shen03}.
In the intermediate time region $750 < t < 1200$ a.u., the NOA is due to a combination of the Kerr effect and 
freed electrons.

In Fig. \ref{fig:long}(b), the pump pulse duration is shortened to 8 fs and the intensity increased to 
$2\cdot 10^{14}$ W/cm$^2$.
The higher intensity induces a NOA from nearly the beginning of the simulation that is a combination of the Kerr 
effect and freed electrons, so that the birefringence due to the Kerr effect alone is no longer dominant.
After the pump pulse has passed, a NOA due to freed electrons alone is observed again, but for times
$t>1500$ a.u. the response becomes isotropic.
Averaging over $1200 < t < 2300$ a.u., $B_r\simeq 0.92$.
This behavior is consistent with the results of Fig. \ref{fig:anisotropy}, where the birefringence weakens for
higher pump values.

\section{Conclusion}

In summary, we have presented computational results based on 3D time-domain Schr\" odinger equation simulations to 
demonstrate that the pump-probe response of a single atom displays a sizeable nonlinear optical anisotropy (NOA)
due to freed electrons for parameters compatible with those in light filamentation in atmospheric gases. 
The origin of the effect is in the anisotropic momentum distribution and spatial correlation in electrons freed by a 
strong linearly polarized pump. 
We found a NOA clearly present for two pump models, namely a delta-function pulse, as well as a more realistic 
optical pump pulse of finite duration.   
Due to numerical limitations, we can only detect a NOA up to times of a few tens of femtosecond after excitation. 
However, it is conceivable that it can persist longer, being bounded from above by the inverse plasma frequency 
for a gas of atoms~\cite{pasMolKoc2011}. 
The NOA revealed here could play a major role in the dynamics of vector light filamentation as the freed
electron contribution will modify the polarization of the propagating filament in a manner not previously considered.

\section*{Acknowledgments}
We thank B. Pasenow, J.~V. Moloney, S.~W. Koch, and G.~I. Stegeman for useful discussions.
This work was supported by the US Air Force Office for Scientific Research under Grant No. FA9550-11-1-0144,
and FA9550-10-1-0561.
Computer time from the University of Arizona High Performance and High Throughput Computing systems 
is gratefully acknowledged.


\end{document}